\documentclass[12pt]{article}
\begin{document}
\begin{large}
\title{\bf {A classical ontology for quantum phenomena.}}
\end{large}
\author{Bruno Galvan \footnote{E-Mail: bgalvan@delta.it}\\ \small Loc. Melta 40, 38014 Trento, Italy.}
\date{\small September 1998}
\maketitle
\begin{abstract}
Quantum mechanics states that a particle emitted at point $(x_1,t_1)$ and detected at point $(x_2,t_2)$ does not travel along a definite path between the two points. This conclusion arises essentially from the analysis of the two-slit experiment, which implicitly assumes (as in the demonstration of the EPR paradox) that a property we will call {\it Independence Property} holds. This paper shows that this assumption is not indispensable. Abandoning the assumption allows to develop an ontology where particle motion is described by classical paths and quantum phenomena are interpreted as a manifestation of a {\it contingent law}, i.e., of a law deriving from the boundary conditions of the universe, such as the second law of thermodynamics. The paper also proposes an equation having a typical quantum-like structure to represent the contingent laws of the universe.
\end{abstract}

\section{Introduction.}

Quantum mechanics states that a particle emitted at time $t_1$ by a source located at $x_1$ and detected at time $t_2$ by a detector located at $x_2$ does not travel along a definite path between the points $(x_1,t_1)$ and $(x_2,t_2)$ and hence does not even travel along the least-action path connecting the two points. This conclusion arises essentially from the analysis of interference phenomena, the most classical of which is the two-slit experiment.

In the analysis of this experiment (as in the demonstration of the EPR paradox) one implicitly assumes that what we will call {\it Independence Property} (IP) is valid. In this paper the IP is formulated precisely and discussed in the framework of a classical dynamical theory, showing that its violation does not entail introducing obscure or unaccountable elements in the theory. The possibility of violating the IP makes it possible to develop an ontology in which particle motion is described by classical paths and quantum phenomena are interpreted as the manifestation of a {\it contingent law}, i.e., of a law deriving from the boundary conditions of the universe, such as for instance the second law of thermodynamics. This paper furthermore proposes an equation having a typical quantum-like structure to represent the contingent laws of the universe.

The IP and its possible violation have been extensively discussed by Price [1,2,3,4,5]. Several concepts presented here have been developed in greater detail in \cite{galvan}.

\section{The Independence Property.}

An experiment on a physical system consists of three phases: (i) the {\it preparation phase}, where the system undergoes a series of manipulations which bring it to the desired initial state; (ii) the {\it execution phase}, where the physical system is allowed to evolve under the influence of controlled environmental conditions; (iii) the {\it measurement phase}, i.e., the interaction of the physical system with a measuring device that records and logs the characteristics of the interaction.

The fundamental feature of an experiment is its reproducibility, i.e., the possibility to provide a description of all the above three phases which is accurate enough to allow another experimenter to perform the experiment and obtain the same outcome. This description will therefore include all the {\it relevant conditions} of the experiment.

Actually, in experimental practice very few experiments happen to be reproducible in this strict sense. This can be explained within a deterministic universe as well, by the fact that in practice it is impossible to prepare the initial state of the system or to set the environmental conditions precisely enough to obtain the same result at every repetition. This problem can be easily solved by taking a sequence of N experiments performed on a single system --which we will call elementary experiments-- as a single experiment, and by taking the statistical distribution of the N respective outcomes as the experimental outcome. The overall experiment is therefore reproducible if the statistical distribution is the same in all its repetitions.

Accordingly, the IP can be stated in the following way:
\begin{quote}
{\bf Independence Property}: {\it in a sequence of N elementary experiments, the statistical distribution of the initial states of the system --i.e., of the states of the system at the end of the preparation phase-- depends only on the preparation phase and not on the execution phase or on the measurement phase.}
\end{quote}
More precisely, if one considers two sequences of two elementary experiments with the same preparation phase, but with different execution and measurement phases, the IP states that the statistical distributions of the initial states of the system are essentially the same for both sequences.

\section{Where the Independence Property is assumed.}

The points where the validity of the IP is assumed to hold in the analysis of the two-slit experiment and in the demonstration of the EPR paradox are now shown.

Reference is made to Heisenberg \cite{heisenberg} for the analysis of the two-slit experiment:

\begin{quote} \small
Quite generally there is no way of describing what happens between two consecutive observations. It is of course tempting to say that the electron must have been somewhere between the two observations and that therefore the electron must have described some kind of path or orbit even if it may be impossible to know which path. This would be a reasonable argument in classical physics. But in quantum physics it would be a misuse of the language which, as we will see later, cannot be justified.
\end{quote}

In order to support this argument, Heisenberg uses the two-slit experiment:

\begin{quote} \small
We assume that a small source of monochromatic light radiates toward a black screen with two small holes in it. [...] At some distance behind the screen a photographic plate registers the incident light. [...]

\small The blackening of the photographic plate is a quantum process, a chemical reaction produced by single light quanta. Therefore, it must also be possible to describe the experiment in terms of light quanta. If it would be permissible to say what happens to the single light quantum between its emission from the light source and its absorption in the photographic plate, one could argue as follows: The single light quantum can come through the first hole or through the second one. If it goes through the first hole and is scattered there, its probability for being absorbed at a certain point of the photographic plate cannot depend upon whether the second hole is closed or open. The probability distribution on the plate will be the same as if only the first hole was open. If the experiment is repeated many times and one takes together all cases in which the light quantum has gone through the first hole, the blackening on the plate will be the same as if only the first hole was open. If the experiment is repeated many times, and one takes together all cases in which the light quantum has gone through the first hole, the blackening of the plate due to these cases will correspond to this probability distribution. If one considers only those light quanta that go through the second hole, the blackening should correspond to a probability distribution derived from the assumption that only the second hole is open. The total blackening, therefore, should just be the sum of the blackenings in the two cases; in other words, there should be no interference pattern. But we know this is not correct, and the experiment will show the interference pattern. Therefore, the statement that any light quantum must have gone {\it either} through the first {\it or} through the second hole is problematic and leads to contradictions. This example shows clearly that the concept of probability function does not allow a description of what happens between two observations. Any attempt to find such a description would lead to contradictions; this must mean that the term `happens' is restricted to the observation.
\end{quote}

It is straightforward to find that the point where Heisenberg assumes that the IP holds is the following sentence: {\it ``If it [the light quantum] goes through the first hole, and is scattered there, its probability for being absorbed at a certain point of the photographic plate cannot depend upon whether the second hole is closed or open."} This sentence in fact implies that if two experiments with the emission of N photons each are performed, the first by opening a single hole in the screen and the second by opening two holes, the number of photons passing through the first hole and being then absorbed in a given point of the plate will be the same in both experiments. If the photons travel along classical paths, the point where the photon is absorbed by the plate depends on the direction of the momentum it has when emitted by the source: the smallest momentum variation will cause a different interaction with the hole and lead therefore to a different impact point on the plate. Therefore Heisenberg assumes that the distribution of initial photon momenta is the same in both experiments although they have different execution phases; in other words, he assumes that IP holds.

However, if both classical paths and IP were to hold one would obtain a result in contrast with experimental data. Accordingly Heisenberg renounces classical paths without even considering the possibility of renouncing the IP.
\vspace{5mm}

Moving on to the EPR paradox, consider its classical demonstration by Bell \cite{bell}: a source emits pairs of particles in opposite directions toward two measuring devices, each of which can measure two distinct observables. Four distinct measurements are therefore possible on each particle pair. The demonstration considers four distinct experiments: in each experiment, one of the four measurements is performed on N pairs. One has, therefore, four experiments with the same preparation phase but with different measurement phases. According to the well-known criteria of local realism, one infers that each particle pair, at the time it is emitted by the source, can be assigned a hidden variable $\lambda$ which univocally determines the result of each one of the four measurements. Let $\rho_i(\lambda), i=1,...,4,$ be the distributions of the variable $\lambda$ in the four experiments. Bell derives his inequality by assuming that the distributions $\rho_i(\lambda)$ are equal in the four experiments; i.e., by assuming that the IP holds. It is trivial to demonstrate that if the distributions $\rho_i(\lambda)$ can be different in the four experiments, Bell's inequality can no longer be demonstrated.

This result is mentioned here merely to show that the IP is repeatedly assumed to hold in the analysis of quantum phenomena; it will not be investigated further.

\section{Dynamical properties and contingent properties.}

As shown, Heisenberg renounces classical paths in order to keep the IP. It is therefore logical to study this property in a classical dynamical framework. In particular --for the sake of simplicity-- this paper considers the universe as a N-particle Hamiltonian system, so that a path of the universe is a curve in the phase-space $R^{6N}$ satisfying Hamilton's equations (or also the least-action principle).

The study of the IP begins by making a distinction between dynamical and contingent properties of a path: a property is termed {\it dynamical} when it derives from the fact that the path satisfies Hamilton's equations and is termed {\it contingent} when it derives from the particular choice of the boundary conditions of the path \footnote{Price calls these two kinds of property {\it law-like} and {\it fact-like} \cite{price3}.}. Clearly a dynamical property holds for all the paths of the universe, whilst a contingent property holds only for some of them. Therefore, in order to prove that a property is contingent it is sufficient to show that it is violated by at least one path.

For example, the energy conservation and angular momentum conservation laws are dynamical properties. The second law of thermodynamics is, on the contrary, a contingent property, because if it holds for a given path than it certainly does not hold for the temporally reversed path, which however is still a dynamically allowed path. Classical statistical regularities --such as the fact that by tossing a coin many times one obtains heads with a 50\% frequency-- also are contingent properties. It is in fact reasonable to admit, for example, that particular paths of the universe exist in which all the tosses of a coin give heads as their outcome. The fact that the statistical regularities of the evolution are contingent properties --i.e., that they cannot be exclusively derived from dynamics-- has been pointed out by Land{\'e} \cite{lande} and Popper \cite{popper}, although these two authors use this result to criticize determinism.

A very important feature of contingent properties is that in general they can predict {\it pre-interactive} or {\it non-interactive} correlations. For instance, there certainly is a path of the universe in which the outcome of the toss of a coin is always heads if the coin is tossed by a blue-eyed person. This is a contingent property, and clearly the blue eyes/heads correlation does not derive from a dynamical interaction between the eyes and the coin but simply by the extremely particular choice of the boundary conditions for the path.

\section{The nature of the Independence Property.}

Two arguments supporting the fact that the IP is a contingent property are now presented.

The first argument is based on the possibility for contingent properties to predict non-interactive correlations. As in the blue-eyes example, it is reasonable to admit that there exist paths which, in a series of experiments, exhibit correlations between the statistical distribution of the initial states of the system and the type of measurement performed on the system, thus violating the IP.

The second argument relies on the fact that the IP is asymmetrical with respect to time. If stated with other words, the IP essentially claims that the state of a system at a given time depends on the interactions experienced by the system in the past but not on the interactions that the system will undergo in the future. This property --like the second law of thermodynamics-- is time-asymmetrical and therefore contingent. For a discussion on the relation between the IP and the second law, see Price [2,3].

Assume, therefore, that the IP is a contingent property.

\section{Is violation of the Independence Property admissible?}

Since the IP is a contingent property, its violation is perfectly admissible from the dynamical point of view and does not require the existence of strange retroactive interactions.

As regards experimental evidence, the IP is certainly verified in macroscopic phenomena. At the microscopic level, however, experimental evidence of its violation is offered by the two-slit experiment itself, if interpreted adequately. In fact, if classical paths are not abandoned, one has to infer that the distribution of the initial momenta of the photons changes according to whether there is a single hole or there are two holes in the screen and that therefore the IP is violated.

Assume, therefore, with Price, that the IP can be violated at the microscopic level.

\section{Are classical paths and quantum phenomena compatible?}

If one allows violation of the IP, the two-slit experiment ceases to be an argument against the fact that particles can travel along classical paths.

Besides the two-slit experiment, there are other quantum phenomena, such as the tunnel effect or quantum diffusion phenomena, that might appear to be incompatible with classical paths \footnote{Actually, it has already been noted that the tunnel effect is not evidence against classical paths \cite{tunnel}.}. However, these phenomena have never been analyzed in depth in this sense. It is furthermore important to clarify --also in view of the above considerations-- the true meaning of demonstrating that a phenomenon is incompatible with classical paths. For example, it is not sufficient to demonstrate that the wave-function of the quantum system is non-vanishing in classically forbidden regions as well. On the contrary, it is necessary to consider a real experiment and to show that none of the classical paths of the universe is compatible with the macroscopic evolution of the experimental apparatus measuring the phenomenon, without excluding paths because they exhibit ``strange" correlations between the apparatus and the quantum system. In other words, it is necessary to prove absolute incompatibility between classical paths and our true perception of the phenomenon, not incompatibility between classical paths and the representation of the phenomenon given by ortodox quantum mechanics, maybe with the addition of constraints such as the requirement that IP must hold.

In my opinion, in this sense no irrefutable proof of such an incompatibility has been provided yet. Therefore, while we await further studies which might refute it, we will formulate the hypothesis that classical paths and quantum phenomena are compatible in the sense explained above.

There are other quantum phenomena, such as photon emission and absorption, particle decay, spin, and so on, that do not appear to be incompatible with classical paths but certainly require an extension of the dynamical framework of classical mechanics. The direction in which this extension might occur is shown in \cite{galvan}.  Without delving into this issue, I merely mention two relevant facts implied by this extension: (i) Hamilton's equations no longer act as dynamical laws and must be replaced by the least-action principle; (ii) the theory loses its deterministic character.

\section{Quantum phenomena like contingent laws.}

If particles travel along classical paths, the interference patterns of the two-slit experiment derive from the particular boundary conditions of the path along which the universe is traveling; accordingly, they are contingent properties. It is natural to extend this conclusion to all quantum phenomena having a statistical nature. Quantum statistical regularities are, therefore, another class of contingent properties which are valid in nature, along with the already-mentioned classical statistical regularities and second law of thermodynamics. We will call {\it contingent laws} the contingent properties that hold in nature.

The possibility that the boundary conditions of the universe might be so particular to give rise to interference patterns is likely to be seen by many physicists as an unacceptable ``conspiracy" of boundary conditions. I believe that this psychological attitude arises mainly from the fact that when speaking of boundary conditions of the universe, one intuitively thinks of the initial conditions. Actually, classical paths, besides initial conditions, also allow boundary conditions associated with the least-action principle, i.e., the initial and final positions of particles. The following paragraph shows that by using this kind of boundary condition it is possible to produce a simple and plausible mathematical equation representing the contingent laws of the universe, including quantum phenomena.
 
\section{Mathematical formulation of contingent laws.}

Consider the contingent law stating that if one tosses a single coin many times one obtains a 50\% frequency of heads. The intuitive explanation of this law is the following: there are ``many more" paths of the universe that have this property than paths that do not, and since the path of our universe was chosen at random, it is very likely that it is among the paths that have this property. Since the sets of paths we are speaking of are sets consisting of infinite elements, in order to say that in a set there are ``many more" paths than in another set we need to define a measure on the set of paths, by which one set has ``more" paths than another set if that measure is greater. The tool by which we will formulate the contingent laws mathematically is, therefore, a measure on the set of paths of the universe. Given this measure, one can say that a contingent property is a contingent law if the measure of the set of paths not having that property is extremely small, vanishing in the limit.

In order to define this measure, it is noted that a measure $\mu$ on the set of paths can be induced by a measure $\mu_B$ on the set of boundary conditions by letting $\mu(\Delta)\equiv\mu_B(\Delta_B)$, where $\Delta$ is a set of paths and $\Delta_B$ is the corresponding set of boundary conditions. For classical paths there are two kinds of boundary condition: (a) the momentum and position of the particles at the initial time $t$, from which one obtains the path in the subsequent times by applying Hamilton's equations; and (b) the positions of the particles at two times $t_1$ and $t_2$, from which one obtains the path between the two times by applying the least-action principle. In order to be able to appropriately deal with the boundary conditions of the universe, assume that the universe was born from a big-bang at the initial time $t=0$, when all particles were spatially concentrated at the point $x=0$. In order to be able to use type (b) boundary conditions, the universe also needs to have a final time $T$; this is not a problem, since as shown in \cite{galvan}, one can then take the limit for $T\rightarrow\infty$ in a natural way. Since the initial positions are set by the big-bang, the type (a) boundary conditions consist of the initial momenta of the particles, and the type (b) boundary conditions consist of the positions of the particles at time $T$.

For type (a) conditions there is a natural measure: volume. If $\Delta_a$ is a set of initial momenta one can set:
\begin{equation}
\mu_C(\Delta_a)\equiv vol(\Delta_a)
\end{equation}
This measure, which we will call {\it classical measure}, considers the initial particle momenta as being independent and uncorrelated and therefore corresponds to the so-called {\it molecular chaos}. Furthermore, it is analogous to the hypothesis of {\it equal a priori probability}, used in statistical mechanics. It is therefore possible for classical statistical regularities and for the second law of thermodynamics to derive from this measure, but it appears rather unlikely that quantum statistical regularities also could derive from it.

A measure with more quantum-like features can instead be defined on type (b) boundary conditions. Let $\Delta_b$ be a set of final positions, and let:
\begin{equation}
\mu_Q(\Delta_b)\equiv \langle 0|e^{i\hat{H}T/\hbar}\hat{\Delta}_be^{-i\hat{H}T/\hbar } |0\rangle,
\end{equation}
where $|0\rangle$ is the improper eigenvector of eigenvalue 0 of the position operator in $L^2(R^{3N})$, $\hat{H}$ is the Hamiltonian operator and $\hat{\Delta}_b$ is the space projector over the set $\Delta_b$. A ``quantum" definition of the  measure $\mu_Q$ is the following: if one considers the universe as a quantum system of N particles that at time $t=0$ lies in the state $|0\rangle$, the measure $\mu_Q(\Delta_b)$ is the (unnormalized) probability of finding the universe in $\Delta_b$ by performing a position measurement at time $T$.

We will call measure $\mu_Q$ {\it quantum measure.} Paper \cite{galvan} shows that in semiclassical approximation the quantum measure is very similar to the classical measure and differs from it by the presence of an interference term.

Therefore we formulate the hypothesis that the quantum measure represents the contingent laws of our universe.

\section{Conclusion.}

The ontology exposed in this paper implies that particles travel along definite paths obeying to two distinct kinds of law: dynamical laws, which derive from the least-action principle; and contingent laws, which derive from the quantum measure $\mu_Q$. Quantum phenomena are a manifestation of a contingent law, and so are classical statistical regularities and the second law of thermodynamics.

We would like to stress that we chose the least-action principle as the origin of dynamical laws rather than Hamilton's equations. This choice arises from the fact that the least-action principle is more general (as noted in section 7) and provides the correct boundary conditions to which the quantum measure is to be applied.

This ontology avoids the known problems of the orthodox interpretation of quantum mechanics: (i) it avoids the conceptual difficulties of wave-particle dualism, since it considers particles exclusively as particles and explains interference phenomena as the manifestation of a contingent law; (ii) it avoids the problem of wave-function reduction, since in this ontology the wave function does not exist; (iii) it has no need to justify the emergence of an approximately classical universe from an underlying quantum one, since it considers the universe to be dynamically classical even at a microscopic level; finally, (iv) it explains the intuitively paradoxical aspects of quantum phenomena, such as the EPR paradox and delayed-choice experiments, by the fact that contingent laws can exhibit pre-interactive or non-interactive correlations.

This paper has formulated two fundamental hypotheses: (i) experimentally observed quantum phenomena are compatible with classical paths, regardless of whether this entails introducing ``strange" correlations between quantum system and measuring devices; and (ii) the quantum measure correctly expresses the contingent laws of the universe. Further studies are needed to support or refute these hypotheses.

\end{document}